\documentclass[sigconf]{acmart}
\AtBeginDocument{%
  }

\setcopyright{acmlicensed}
\copyrightyear{2026}
\acmYear{2025}
\acmDOI{XXXXXXX.XXXXXXX}
\acmConference[CHI '26]{Proceedings of the 2026 CHI Conference on Human Factors in Computing Systems}{April 13--17, 2026}{Barcelona, Spain}

\usepackage{booktabs, makecell, array}

\usepackage{graphicx}
\usepackage{subcaption}

\usepackage{amsmath}

\usepackage{listings}
\usepackage{color}

\definecolor{dkgreen}{rgb}{0,0.6,0}
\definecolor{gray}{rgb}{0.5,0.5,0.5}
\definecolor{mauve}{rgb}{0.58,0,0.82}

\begin{document}

\title{World Mouse: Exploring Interactions with a Cross-Reality Cursor}

\author{Esen K. Tütüncü}
\authornote{Both authors contributed equally to this research.}

\orcid{0000-0002-0050-0908}
\affiliation{%
  \institution{Institute of Neurosciences\\ of the University of Barcelona}
  \city{Barcelona}
  \country{Spain}}

\author{Mar Gonzalez-Franco}
\authornotemark[1]
\affiliation{%
  \institution{Google}
  \city{Seattle}
  \country{USA}}
  
\author{Khushman Patel}
\affiliation{%
  \institution{Google}
  \city{Mountain View}
  \country{USA}}

\author{Eric J. Gonzalez}
\affiliation{%
  \institution{Google}
  \city{Seattle}
  \country{USA}}

\renewcommand{\shortauthors}{K. Tütüncü, Gonzalez-Franco et al.}

\begin{abstract}

As Extended Reality (XR) systems increasingly map and understand the physical world, interacting with these blended representations remains challenging. The current push for ``natural'' inputs has its trade-offs: touch is limited by human reach and fatigue, while gaze often lacks the precision for fine interaction. To bridge this gap, we introduce \textit{World Mouse}, a cross-reality cursor that reinterprets the familiar 2D desktop mouse for complex 3D scenes. The system is driven by two core mechanisms: within-object interaction, which uses surface normals for precise cursor placement, and between-object navigation, which leverages interpolation to  traverse empty space. Unlike previous virtual-only approaches, World Mouse leverages semantic segmentation and mesh reconstruction to treat physical objects as interactive surfaces. Through a series of prototypes—including object manipulation and screen-to-world transitions—we illustrate how cross-reality cursors may enable seamless interactions across real and virtual environments. 

\end{abstract}


\begin{teaserfigure}
\centering
    \includegraphics[width=1\textwidth]{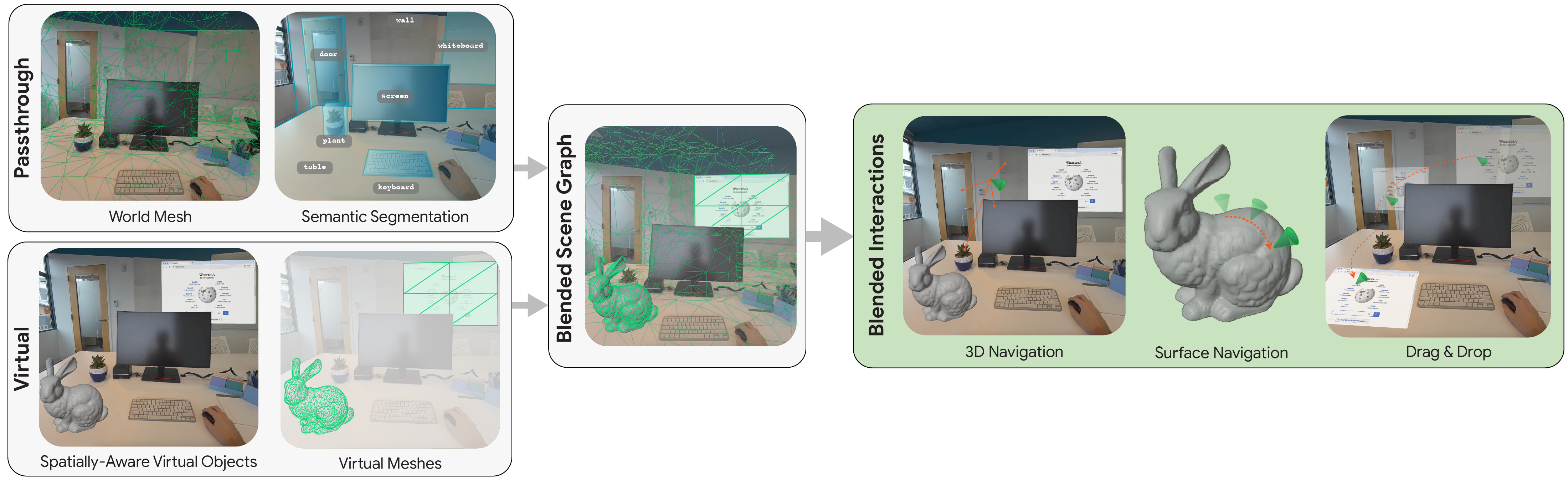}
    \caption{\textit{World Mouse} combines virtual content with a semantically labeled real-world mesh, enabling seamless cross-reality cursor-based interactions such as: 3D selection and manipulation, object surface navigation, and drag-and-drop.}
  \label{fig:teaser}
\end{teaserfigure}



\maketitle

\section{Introduction}

While XR interactions often emphasize hand ray or mid-air gestures \cite{argelaguet2013survey,wentzel2024comparison}, these techniques can induce fatigue over time and lack the subtle control experienced mouse users rely on \cite{restfulraycast, handover}. This is partially due to XR's heavy reliance on direct inputs \cite{gonzalez2024guidelines}. Even more advanced hybrid inputs such as gaze+pinch \cite{pfeuffer2017gaze+} still demand physical effort through bimanual interactions \cite{handson}, preshaping \cite{mikkelsen2026preshaping} or mid-air gesturing  \cite{pinchcatcher,forcepinch}. In contrast, the mouse’s traditional strengths remain compelling \cite{engelbart1968research}:  many immersive applications would benefit from precise, indirect input--especially those involving long work sessions or detailed tasks like 3D modeling and data annotation \cite{hansberger2017dispelling, sun2018comparing, SymbiosisSketch, handover}. Moreover, since many XR hardware setups already permit seated or desk-based use, the mouse is well-positioned to serve as a high-performance spatial input, provided the complexity of 3D depth can be addressed. Towards this, previous research has explored mapping the desktop and mouse to the virtual environment \cite{chen2024luxar,fender2017meetalive} or directly integrating the mouse to virtual modeling \cite{zhou2022depth, SymbiosisSketch}.


We advance cursor-based XR interaction by adapting the desktop mouse for blended cross-reality environments (i.e., containing both real and virtual interactable elements, see Figure \ref{fig:teaser}). Our approach, which we call the \emph{World Mouse}, leverages a blended scene graph to map 2D mouse movements into 3D interactions across spatially distributed objects, environments, and interfaces. We build upon two core mechanisms. This approach rely on two core mechanisms: First, \textit{within-object interaction}
follows the logic of rasterization to precisely track the mouse's 2D movements across the 3D surface of a single object \cite{midair21}. Second, \textit{between-object navigation} enables fluid movement across disjoint elements by leveraging an ``invisible mesh'' that interpolates between  nearby objects \cite{armeni20193d}. Together, these techniques allow users to effortlessly navigate objects of varying scales and depths, mirroring the simplicity of a desktop cursor but applied to the physical world.

In this paper, we outline the conceptual foundation and technical implementation of the World Mouse, detailing how it dynamically infers and adapts cursor depth in real time. Through a series of prototypes, we demonstrate how this approach provides the precise, familiar control of a desktop mouse while avoiding the fatigue of mid-air gestures. Beyond purely virtual environments, we show how the World Mouse integrates with passthrough augmented reality, enabling seamless interactions across both digital and physical objects in a blended scene.. Ultimately, by augmenting a universally understood device with robust 3D capabilities, we expand the reach of traditional workflows and invite a reimagined role for classic 2D input in spatial computing.

\section{Background \& Related Work}
We contextualize our work by highlighting three key trajectories in spatial input: the evolution of indirect precision input, the limitations of embodied XR techniques, and recent efforts to extend cursor-based interaction into immersive systems.

\subsection{Precision in HCI}

The pursuit of precision through indirect input underpinned much of early human-computer interaction \cite{hinckley2007input}. The adoption of the mouse in early graphical operating systems of course marked a major leap in usability \cite{engelbart1968research}. Simultaneously, early explorations into ``natural'' interactions involving gestures and voice emerged \cite{bolt1980put}, even though the hardware of the time could not fully support researchers' vision. The advancement of computing and sensing capabilities (e.g. via commodity IMUs), however, gradually laid the groundwork for modern motion-based interfaces \cite{hinckley2000sensing}.



As direct touch interactions became ubiquitous in the 2000s \cite{hinckley1998two}, they introduced a fundamental tradeoff: while touch fosters immediate and intuitive actions, it inherently sacrifices precision \cite{holz2011understanding}. Although styli can restore accuracy \cite{xia2015nanostylus}, pen-based input has largely remained a niche solution. As digital displays scaled up, novel solutions addressed the challenge of distant interaction, ranging from the
explicit transitions between direct and relative input, as in \textit{HybridPointing} \cite{forlines2006hybridpointing}, to pen+hand manipulations for massive displays like \textit{WritLarge} \cite{xia2017writlarge}. This historical near-versus-far dilemma mirrors current challenges in spatial computing, where interaction is typically framed as either direct touch or distant raycasting \cite{gonzalez2024guidelines}.

\subsection{XR Interaction Techniques}


XR interaction has evolved from early controller-based inputs \cite{anthes2016state} to incorporate full hand and body tracking \cite{gonzalez2020movebox}, alongside more recent voice and gaze-based techniques \cite{ pfeuffer2024design}. Historically, immersive experiences favored near-field interactions (0.5–3.5 meters) to simplify depth perception and minimize errors \cite{zhou2023design}. However, as spatial computing advances, there is a need for adaptable input techniques capable of handling diverse interaction ranges.


Still, embodied mid-air input can cause user fatigue \cite{hansberger2017dispelling, palmeira2023quantifying} and may lack the accuracy required for delicate tasks. To alleviate this, researchers have explored clutching or altering control-display (CD) gains in XR, such as in \textit{Go-Go} \cite{poupyrev1996go}, \textit{Erg-O} \cite{montano2017erg}, and \textit{SnapMove} \cite{cohn2020snapmove}. Extensive surveys and evaluations of 3D object selection techniques \cite{argelaguet2013survey,lee2003evaluation} reveal that few incorporate adaptive pointing \cite{konig2009adaptive} or special CD gains, such as \textit{PRISM} \cite{frees2007prism}. Critically, most systems still heavily rely on raycasting; a study of consumer VR games found that approximately 60\% use controller-based rays for selection, vastly outpacing direct hand interaction \cite{wentzel2024comparison}.


To improve far-distance accuracy and reduce full-arm exertion, recent work has explored micro-gestures \cite{pei2024ui}, gaze-plus-pinch techniques \cite{pfeuffer2024design}, and pinpointing \cite{kyto2018pinpointing}. Other research circumvents the problem by repurposing everyday devices--such as smartphones \cite{zhu2024phoneinvr, zhu2020bishare} or headphones \cite{panda2023beyond}--as convenient input proxies. Supported by cross-device toolkits like XDTK \cite{gonzalez2024xdtk}, this also facilitates a shift toward multi-device experiences where users can interact with shared spatial content even without dedicated XR hardware \cite{kitson2024virtual}.

\begin{figure*}[h]
    \centering
    \includegraphics[width=1\linewidth]{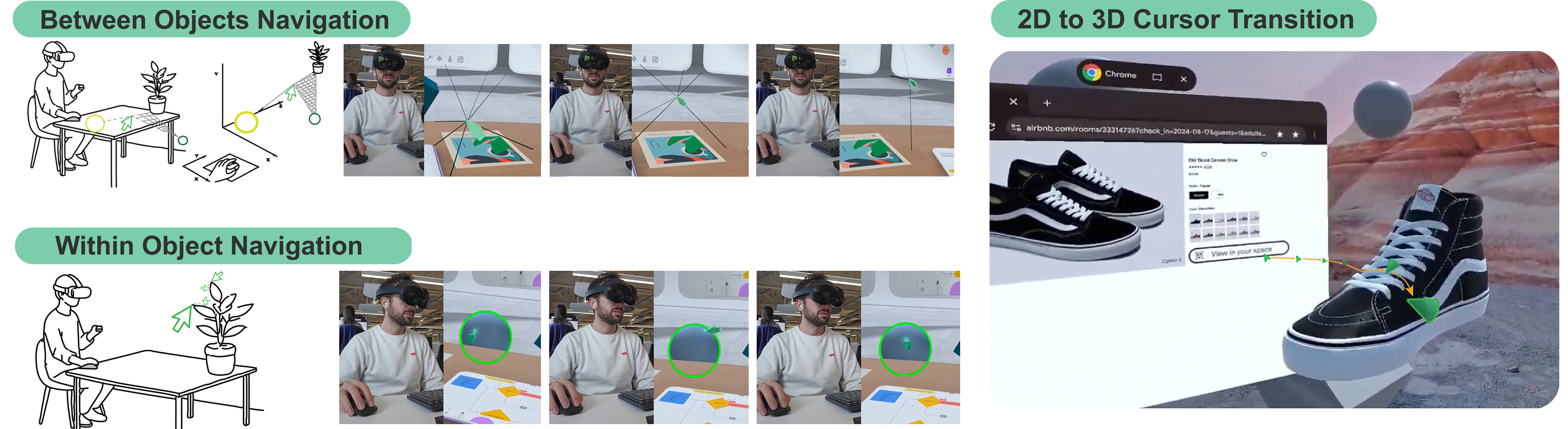}
    \caption{Cursor behaviors explored through the World Mouse. Left (Upper): Between Objects Navigation leverages interpolation allowing the cursor to traverse through the space between objects. Left (Lower): Within Object Navigation allows the cursor to move along the surface of any object. Right: 2D to 3D Cursor Transition enables a smooth visual transition when using moving from 2D panels like browsers to a 3D objects.}
    \label{fig:2dmouse3d}
\end{figure*}

\subsection{Cursor-based Interactions in XR}

Efforts to enhance precision and reduce fatigue have prompted researchers to revisit cursor-based methods for immersive systems. For instance, \textit{Hyve-3D} \cite{dorta2016hyve} introduced a 3D cursor for precise collaborative design, demonstrating that minimizing full-arm extensions improves user performance. Similarly, Sun et al. \cite{sun2018} found that the desktop mouse can outperform hand tracking and trackpads in VR, particularly during seated work on stable surfaces.

Recent work has refined these insights. The \textit{In-Depth Mouse} \cite{zhou2022depth} introduced a depth-adaptive cursor in VR, while the \textit{Everywhere Cursor} \cite{kim2022everywhere} demonstrated that adapting 2D input devices for spatial AR can boost both familiarity and precision. These efforts demonstrate the viability of cursor-based interaction in immersive environments, but they largely address interactions within a single modality (VR or AR). In contrast, we investigate how a depth-adaptive cursor can operate across fully blended environments. Building on these works, the World Mouse introduces an interaction model that seamlessly traverses virtual content and the physical world, supporting flexible, cross-reality workflows.

\section{World Mouse Design and Implementation}

The World Mouse is an indirect input method designed to provide high precision when interacting with both real and virtual objects. By mapping the standard 2D computer mouse into a blended 3D environment, the system allows users to manipulate spatial assets, navigate complex scenes, and seamlessly interact with digital and analog elements alike.

\subsection{Bringing the 2D Mouse into 3D Space}




We build directly upon the Depth-Adaptive Cursor introduced by the \textit{In-Depth Mouse} \cite{zhou2022depth}, which enables mouse interaction in VR by casting a ray from the user's viewpoint and utilizing Voronoi-based interpolation to infer depth in empty space. Specifically, the system maps standard 2D mouse deltas to angular displacements of the cursor ray, converting planar movement into spherical rotation relative to the user's viewpoint. While their approach explores cursor-based manipulation in purely virtual scenes, we extend this logic to fully blended cross-reality environments.

Rather than relying on discrete screen-space segmentation, World Mouse treats the physical and virtual world as a unified, continuous interaction space. By leveraging a geometric reconstruction of the physical environment (e.g., via Meta's Scene API or Android XR's Scene Meshing), we can generate a blended scene graph linking the surfaces of both real-world geometry and virtual assets -- effectively acting as an ``invisible mesh'' connecting the two. This allows the cursor to fluidly traverse the scene, dynamically adjusting its depth based on proximity to physical and digital objects alike. We detail this implementation through three distinct cursor behaviors: navigation within object surfaces, interpolation between objects, and the transition between 2D and 3D spaces.

\subsubsection{Within Object Navigation}

Similar to \textit{In-depth Mouse}, when the cursor intersects with an object, it will traverse on the object's surface. Specifically, while planar mouse movement translates to relative movement of the cursor within the user's field of view, the cursor's depth is adjusted to match the object surface depth at that point. Furthermore, the system dynamically aligns the cursor’s orientation with the local surface normal at the point of contact to provide further feedback about the surface topography. As a result, users can intuitively navigate object surfaces to make precise selections and manipulations (Fig. \ref{fig:2dmouse3d}).


\subsubsection{Between Objects Navigation}

To navigate the voids between objects, the system utilizes the blended scene mesh to establish a continuous interaction surface. Instead of casting a fixed distance into empty space, the cursor moves along an interpolated path driven by the distance to adjacent geometry. This creates a dynamic "bridge" between varying depths, computed in real-time based on the spatial relationships of visible targets. This mechanism allows the cursor to fluidly transition toward destination objects while maintaining precision and stability (Fig. \ref{fig:2dmouse3d}).

\subsubsection{2D to 3D Cursor Transition}
In XR, users may need to repeatedly switch between interactions with 2D windows and 3D content. Previous work such as \textit{HandOver} \cite{handover} has explored the transition between standard 2D mouse usage and 3D interaction. With World Mouse, we extend this to explore how cursors may transition when moving from a 2D interactive panel out into 3D space, including the user's surroundings (Fig.~\ref{fig:2dmouse3d}). For example, a user might finish retouching an image in a 2D editing application and then slide the cursor out of the application window into the 3D canvas to place the edited image as a poster on a physical wall.

\subsection{Bringing the Mouse into the Real World}
To use a 3D mouse in the real world, we need a robust understanding of the real-world geometry such that the mouse can navigate between real-world elements. For true interactivity, this requires object detection and segmentation, with extra consideration for regions that may be relevant for 2D interactables such as virtual windows, texts, and screens. Recent advances in computer vision have improved understanding of objects and 3D world structure, which we can leverage to provide a basic surface for mouse navigation. The system constructs a mesh from the convex hulls of these detected objects relative to the user's camera view. Designed for stability, this surface accommodates both static and dynamic objects without requiring expensive recalculation for every viewpoint shift. This results in a continuous interactable surface, unlike the discrete zones in VR, allowing for precise mouse usage within 2D contexts like physical monitors or virtual desktop windows.

\begin{figure*}[h]
    \centering
    \includegraphics[width=0.9\linewidth]{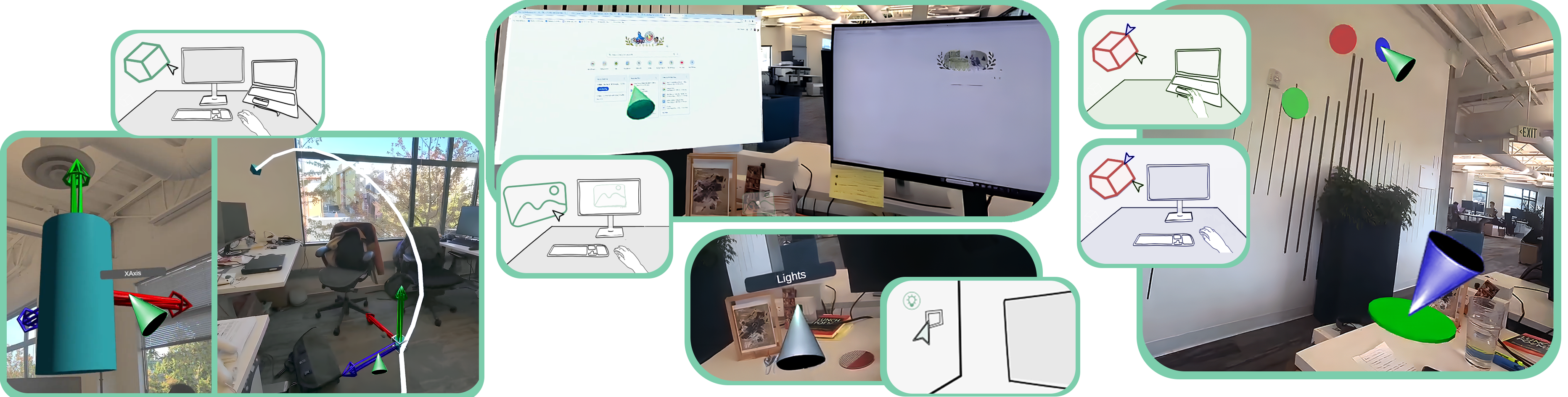}
    \caption{World Mouse Application Scenarios. (Left) Spatial Authoring: Users can spawn virtual objects and manipulate them using high-precision 3D gizmos, including spline editing and vertex-snapping to real-world meshes. (Middle) Cross-Device \& IoT Control: Digital content can be transitioned from physical screens into the environment, while IoT devices are controlled via interactive proxies and re-rendered reality filters. (Right) Social XR: The system supports collaborative workflows allowing multiple users to interact with shared virtual assets.}
    \label{fig:applications}
\end{figure*}

\section{Prototype Explorations}

To demonstrate how the World Mouse translates into tangible, everyday workflows, we developed a series of prototype interactions (Figure \ref{fig:applications}). Each illustrates a different aspect of precision input in a mixed environment, from simple selection and manipulation to cross-device collaboration and IoT control.

\subsection{The Desktop Metaphor in 3D Space}
We first explore how the established metaphors of desktop computing -- including point-and-click selection, hover-state feedback, and context-sensitive menus -- might map to interactions with a blended 3D environment. To support these paradigms, the cursor continuously aligns with environment geometry and provides visual highlights for interactive elements similar to match standard mouse interactions.

\textbf{Spatial Clipboard (Select, Copy, and Paste):} 
Users can click to select virtual objects or areas in the passthrough view and take screenshots of the current augmented reality scene, capturing both virtual and real-world elements. 3D models from the physical world (scanned via the XR headset) can also be pasted it into the virtual environment. This interaction would import real-world objects into the digital space, creating a seamless blend of physical and virtual.

\textbf{Right-Click and Contextual Menus:} 
Right-clicking triggers radial context menus that appear around the 3D cursor. These menus ``fan out'' options such as ``Properties,'' ``Copy,'' or ``Spawn Object,'' with contents that adapt based on the semantic understanding of the target object. This allows users to quickly switch modes or access settings without navigating distant floating UIs.

\textbf{Scroll for Zoom and Depth:} In 3D space, the mouse scroll-wheel can offer valuable continuous input and disambiguation. For example, users can use the scroll-wheel to fluidly manipulate the depth of dragged objects, adjust the scale of virtual objects, or adjust the zoom of magnified views in real space.

\subsection{3D Authoring and Manipulation}
Beyond typical 2D interactions, the World Mouse enables high-precision spatial creation tools that are typically fatigue-inducing when using mid-air gestures alone.

\textbf{Spawning and Anchoring Objects:}
The user can bring up a palette to select objects or 3D widgets to place into the environment. Upon selection, a “ghost” version of the new object attaches to the cursor, letting the user choose exactly where to drop it in physical or virtual space. For example, spawning a virtual note or sticky label to annotate meeting notes on a real white board.

\textbf{3-Axis Gizmo and Drag \& Drop:}
For CAD-like control, a 3D transform gizmo provides $x, y, z$ handles for fine-grained manipulation. Objects can be moved freely through space or assigned mass to follow physics, sliding across surfaces and balancing on physical planes. The precision is high enough to allow spline editing in 3D space or attaching vertices to real-world meshes.

\subsection{Cross-Reality and Ambient Interaction}
The World Mouse can also bridge digital content with physical environments through connected devices and scene understanding.

\textbf{Virtual Window:}
 The 3D cursor can manipulate 2D panels to place them within the real-world mesh. This decoupling from physical monitors allows for expanded, ergonomic viewing angles while maintaining application continuity, regardless of where the compute happens.


\textbf{IoT interactivity:}
By leveraging semantic labels users can assign interactive proxies to physical objects. Hovering or clicking these proxies allows for direct control of real devices, such as toggling lights or adjusting thermostats. Some interactions may trigger a "re-rendered reality," such as using a passthrough filter to simulate turning off lights. 


\subsection{ Cross-Device and Multi-User Reality}

\begin{figure}[h]
    \centering
    \includegraphics[width=\linewidth]{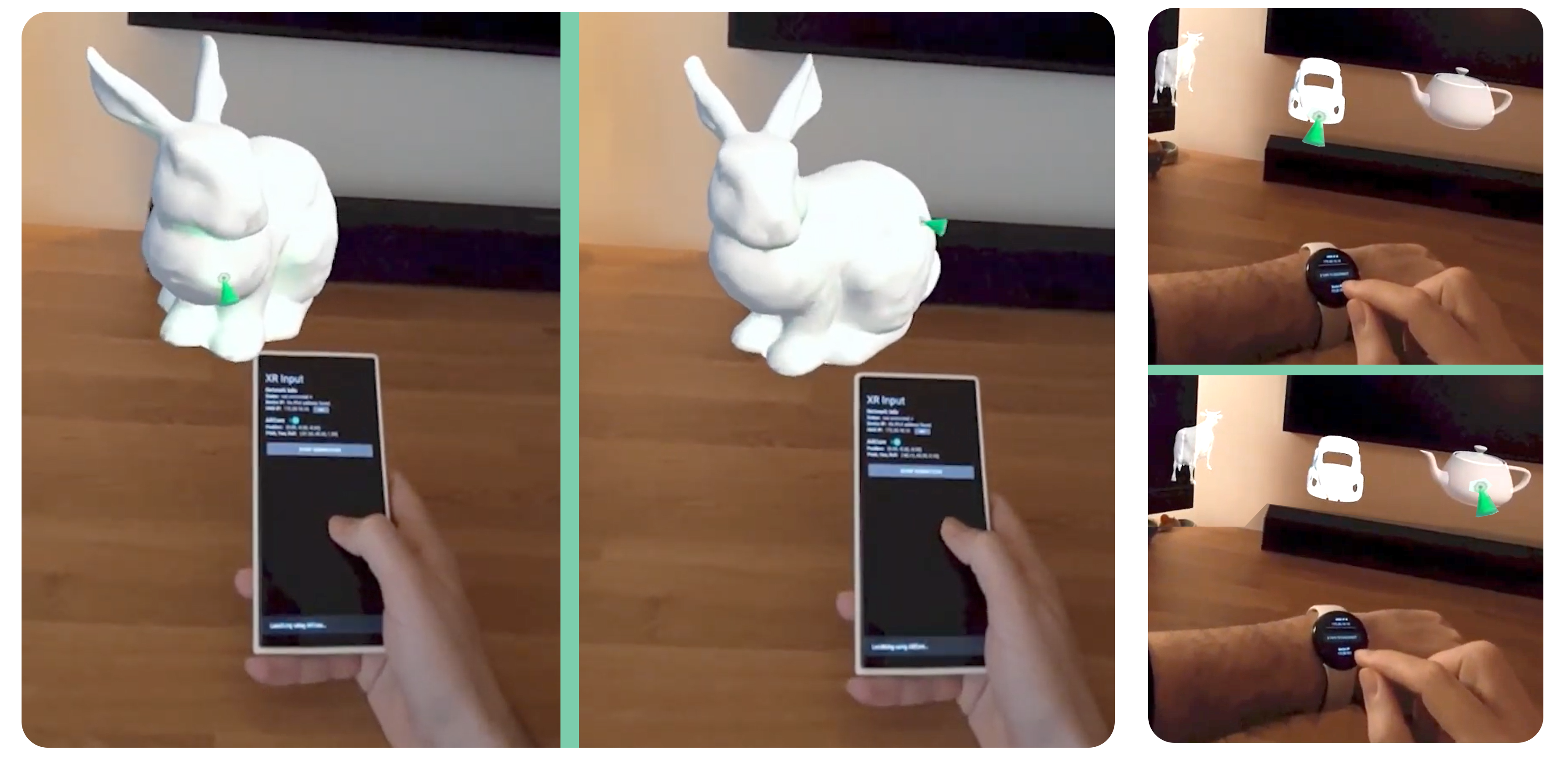}
    \caption{Controlling the World Mouse using the touchscreen of a smartphone (left) or smartwatch (right) via XDTK \cite{gonzalez2024xdtk}.}
    \label{fig:phonemouse}
\end{figure}

The integration of personal devices like smartphones as World Mouse controllers in XR environments offers a compelling solution for high-precision input in XR environments. By using precise touchscreens as trackpads for cursor control, these devices provide a level of accuracy and familiarity that often surpasses contemporary hand-tracking or controller-based raycasting (Fig. \ref{fig:phonemouse}).


Future iterations could support multi-user blended reality, facilitating collaboration across co-located \cite{kitson2024virtual} and remote setups \cite{wang2024avatarpilot,gronbaek2023partially}. Because the cursor interaction model is backward compatible, users on standard laptops or non-immersive displays can also participate by interacting with a 2D projection of the shared scene graph. This enables cross-device engagement comparable to multiplayer gaming \cite{kitson2024virtual} or distributed productivity \cite{bragdon2011code}, perhaps even enhanced by proxemics and not just by cursors \cite{marquardt2012cross,gonzalez2015framework, gronbaek2020proxemics}.

\subsection{Interacting with AI}



Recent advances in depth sensing and semantic segmentation~\cite{tian2024diffuse,kirillov2023segment} have expanded the world-understanding capabilities of XR systems from simple passthrough video feeds to semantically-aware environment reconstructions. This context is essential for interactions with AI agents, which require clear deictic references--knowing exactly which object a user is discussing--to function effectively~\cite{bovo2024embardiment}. While systems like \textit{XR-Objects}~\cite{dogan2024augmented} and \textit{LightAnchors}~\cite{ahuja2019lightanchors} demonstrate the potential of attaching digital information to physical items, relying solely on voice or mid-air gestures to select these targets often leads to ambiguity or fatigue. As a results, we believe precise cursor-based interaction may be a valuable for AI interactions in XR, providing a high-fidelity mechanism to ``ground'' prompts to specific real-world targets without the ambiguity of gaze or the exertion of gesturing.

\section{Discussion}

The World Mouse invites reconsideration of the dominant narrative that spatial computing must discard the desktop mouse in favor of novel or exotic interaction paradigms. Rather than positioning indirect input as obsolete, we believe our explorations suggest that precision and familiarity--hallmarks of the traditional mouse--still have great value when recontextualized for immersive and hybrid environments. By decoupling depth from physical reach, this approach establishes a low-fatigue interaction model that preserves dexterity while avoiding the inaccuracy and fatigue often associated with raycasting.

Moreover, our work illustrates that continuity can be a powerful strategy for designing cross-reality interactions. This interaction technique offers a bridge: it enables users to effortlessly flow between 2D and 3D spaces, to transition across digital and physical layers, and to interact with environments that are simultaneously virtual, real, and shared.

The implications extend beyond usability. As AI systems increasingly permeate our spatial environments by generating content, mediating interfaces, and controlling physical devices, there is a growing demand for tools that can express intent with both clarity and granularity. The World Mouse suggests one possible direction: a precision instrument for spatial design, AI guidance, and cross-device coordination. It can serve not just as an input device, but as a thinking tool in the age of spatial and generative computing.

While the World Mouse is designed to streamline 3D selection and confirmation tasks in XR, it is not intended for unconstrained freehand sketching \cite{midair21}. For scenarios requiring fluid, continuous 3D input (e.g., mid-air drawing or sculpting), traditional freehand methods may be better suited. Nevertheless, if XR becomes an essential interface to AI, much like screens have been for desktop computing, the proposed technique’s strength in precision and reduced fatigue could prove valuable. By allowing users to pinpoint where and how AI-driven tools (both large language models and computer vision systems) should focus, the World Mouse bridges physical and digital domains, supporting a range of emerging spatial workflows. We view this work not as a replacement for embodied interaction, but as a complementary trajectory -- one that foregrounds precision, continuity, and long-term usability in spatial computing.

\section{Conclusion}
In this work, we explored how the familiar form of a desktop mouse can be extended into immersive environments, offering a viable alternative for precise, low-fatigue interaction across varying depths and realities. The World Mouse bridges longstanding input metaphors with the demands of spatial computing, providing users with a consistent and intuitive control mechanism that remains effective even as environments shift between physical and digital. Rather than replacing gestural techniques, our approach contributes to a broader design landscape where diverse input modalities coexist and are tailored to specific user needs. As XR and AI systems continue to evolve, tools like the World Mouse may play a meaningful role in expanding interaction possibilities while preserving user agency.


\bibliographystyle{ACM-Reference-Format}
\bibliography{sample-base}

\end{document}